\documentclass[aps,prd,floatfix,superscriptaddress,twocolumn,showpacs]{revtex4}

\usepackage{epsfig}
\usepackage{amsmath}
\usepackage{graphicx,psfrag}
\usepackage{dcolumn}
\usepackage{bm}
\usepackage{slashed}
\usepackage{color}
\usepackage{soul}

\newcommand{\beq}{\begin{equation}}
\newcommand{\eeq}{\end{equation}}
\newcommand{\bea}{\begin{eqnarray}}
\newcommand{\eea}{\end{eqnarray}}
\newcommand{\hf} {\frac{1}{2}}

\newcommand{\nn}{\nonumber\\}

\newcommand\fig[1]     {Fig.\,{\ref{#1}}}

\def\eq#1{(\ref{#1})}
\def\s0#1#2{\mbox{\small{$ \frac{#1}{#2} $}}}
\def\0#1#2{\frac{#1}{#2}}

\def\mr#1{{\mathrm{#1}}}

\begin{document}

\title{Truncation Effects in the Functional Renormalization Group Study \\ of Spontaneous Symmetry Breaking} 

\author{N. Defenu}
\affiliation{SISSA, Via Bonomea 265, I-34136 Trieste, Italy}
\affiliation{CNR-IOM DEMOCRITOS Simulation Center, Via Bonomea 265, I-34136 Trieste, Italy}

\author{P. Mati}
\affiliation{Institute of Physics, Budapest University of Technologyand Economics, H-1111 Budapest, Hungary}
\affiliation{University of Debrecen, P.O.Box 105, H-4010 Debrecen, Hungary}
\affiliation{Institute of Physics, E\"otv\"os University, H-1117 Budapest, Hungary} 
  
\author{I. G. M\'ari\'an}
\affiliation{University of Debrecen, P.O.Box 105, H-4010 Debrecen, Hungary}

\author{I. N\'andori}
\affiliation{University of Debrecen, P.O.Box 105, H-4010 Debrecen, Hungary}
\affiliation{MTA-DE Particle Physics Research Group, P.O.Box 51, H-4001 Debrecen, Hungary}
\affiliation{MTA Atomki, P.O. Box 51, H-4001 Debrecen, Hungary} 

\author{A. Trombettoni} 
\affiliation{CNR-IOM DEMOCRITOS Simulation Center, Via Bonomea 265, I-34136 Trieste, Italy}
\affiliation{SISSA, Via Bonomea 265, I-34136 Trieste, Italy}
\affiliation{INFN, Sezione di Trieste, Via Bonomea 265, I-34136 Trieste, Italy}

\begin{abstract} 
We study the occurrence of spontaneous symmetry breaking (SSB) for $O(N)$ models 
using functional renormalization group techniques. We show that even the local potential 
approximation (LPA) when treated exactly is sufficient to give qualitatively correct 
results for systems with continuous symmetry, in agreement with the Mermin-Wagner 
theorem and its extension to systems with fractional dimensions. For general $N$ 
(including the Ising model $N=1$) we study the solutions of the LPA equations for 
various truncations around the zero field using a finite number of terms (and different regulators), 
showing that SSB always occurs even where it should not. The SSB is signalled by Wilson-Fisher 
fixed points which for any truncation are shown to stay on the line defined by vanishing 
mass beta functions. 
\end{abstract}


\maketitle

\section{Introduction} 
\label{intro}
Spontaneous symmetry breaking (SSB) is a cornerstone concept in a variety of systems, 
from field theory and particle physics to statistical mechanics and interacting lattice models. 
The study of the occurrence of SSB play a crucial role in the theory of phase transitions 
and in the characterization of ordered phases and it highlights the interplay between SSB 
and the dimensionality of the system: this interplay is customarily expressed by defining a 
lower critical dimension $d_L$ for which SSB cannot occur \cite{Huang87}.  

A celebrated exact result connecting SSB and dimensionality is provided by the 
Mermin-Wagner theorem  \cite{Mermin66,Hohenberg67,Coleman73}. According to the 
Mermin-Wagner theorem a continuous symmetry cannot be spontaneously broken in 
two dimensions: $d_L=2$. This theorem has been formulated for classical systems 
\cite{Mermin66} and then extended to quantum systems \cite{Hohenberg67,Coleman73}. For 
magnetic systems with continuous symmetry it rules out the possibility of having a non-vanishing 
magnetization at finite temperature in two dimensions, and for $2d$ interacting Bose gases 
predicts that no Bose-Einstein condensation occurs at finite temperature \cite{Hohenberg67} 
(for Bose gases this result has been extended to zero temperature \cite{Stringari95}). As well 
known, even though the Mermin-Wagner theorem rules out SSB and the existence of a local 
order parameter in two dimensions, nonetheless the Berezinskii-Kosterlitz-Thouless transition 
may yet occur for the $U(1)$ symmetry and it signaled by the algebraic behaviour of correlation 
functions in the low temperature phase \cite{Kadanoff00}.

The Mermin-Wagner theorem for the $O(N)$-symmetric scalar field theory states that for 
$N \ge 2$ in two dimensions no SSB occurs. Although originally formulated in integer 
dimensions, this result was later extended to graphs with fractional dimensions \cite{Cassi92}: 
in this way one can explicitly show that for $N \ge 2$ there is no SSB for $d \le 2$, with $d$ being 
real, while SSB occurs for $d>2$ \cite{Burioni96}. In \cite{codello:13} the study of how $O(N)$ 
universality classes depend continuously on the dimension $d$ (and as well on $N$), in particular 
for $2<d<3$ was recently presented. The Ising model, i.e. the $N = 1$ case, is different from 
$O(N)$ models with continuous symmetry ($N\ge 2$) because the symmetry is discrete: in two 
dimensions it notoriously has a finite temperature phase transition \cite{Mussardo10} and it 
can be shown that this happens for $d \ge 2$ with $d$ real \cite{Burioni99}. The large 
$N$-limit of $O(N)$ models is also interesting because for $N \to \infty$ it is equivalent to 
the spherical model \cite{Stanley68}, which is exactly solvable \cite{Joyce73}. 

The $O(N)$ model represents then an ideal playground to study the interplay of SSB and 
dimensionality and to test whether (and how) the appearance of SSB depends on the 
approximation schemes. A powerful method used to consider the phase structure of a model, 
and consequently to study the appearance of SSB, is the functional renormalization group 
(FRG) method \cite{Wi1971,rg1,rg2,rg3,rg4,rg5,rg6,rg7}. The $O(N)$ model has extensively 
studied using FRG approaches: as relevant for our purposes, we mention it was used to study as a 
function of dimension critical exponents of $O(N)$ models \cite{Ballhausen03,codello:13,Codello14} 
and to investigate truncation effects and the regulator-dependence of the FRG equation  
\cite{scheme_o(n)_1,pms_o(n),scheme_o(n)_2,bmw_o(n)_1,scheme_o(n)_3,bmw_o(n)_2,scheme_o(n)_4,scheme_o(n)_5}, 
while a FRG study of the critical exponents of the Ising model for $d<2$ was presented 
in \cite{Ballhausen03}. The study of single-particle quantum mechanics can be seen as a 
"low-dimensional'' statistical mechanics model: FRG studies addressed double well potential 
and quantum tunneling \cite{low_d_o(n):1,low_d_o(n):2} and quartic anharmonic oscillators 
\cite{low_d_o(n):3}.

In the FRG framework one has to solve an integro-differential equation valid for functionals 
which is usually handled resorting to approximations. It is in general of great importance to 
know "how good" are the used approximations and to test them against exact results. An 
approximation commonly used is the Local Potential Approximation (LPA), in which the 
wave-function renormalization and higher derivative terms in the effective action are discarded, 
resulting  in a vanishing anomalous dimension \cite{rg1,rg2,rg3,rg4,rg5,rg6,rg7}. Furthermore, 
one often treats LPA introducing further approximations via the introduction of a finite number 
of couplings, defined as the Taylor coefficient of an expansion of the potential around the zero (or
the minimum) of the field, and studying their renormalization group (RG) flow. 

In this paper our goal is two-fold: \textit{i)} from one side we aim at discussing what level of 
approximation is needed to reproduce the Mermin-Wagner theorem and to show that no SSB 
occurs for $d\le2$ with $d$ real and $N\ge2$, \textit{ii)} from the other we intend to investigate 
how truncation affects the occurrence or non-occurrence of SSB comparing/contrasting the 
obtained findings with the exact prediction of  the Mermin-Wagner theorem. Our findings for 
systems with continuous symmetry can be summarized as follows:
\begin{itemize}
\item \textit{i)} LPA, when  treated exactly, is enough to reproduce the Mermin-Wagner theorem; 
\item \textit{ii)} defining the couplings as the coefficients of a Taylor expansion of the effective
potential centered in the zero, we have that,
for any finite number of couplings in LPA, SSB appears also when it should not 
(i.e., for $d\le2$), and the corresponding (spurious) Wilson-Fisher fixed points lie on the line 
defined by vanishing mass beta functions. On the other hand
using a Taylor expansion of the potential around the minimum we can recover Mermin-Wagner theorem
even for finite number of couplings.
\end{itemize}
For the Ising model ($N=1$) the SSB occurs for $d>2$, again in agreement with exact results. 
We will also discuss in detail the limit $N \to \infty$ (the spherical model) where the LPA is exact 
\cite{Tetradis94} (since the anomalous dimension of the spherical model is vanishing) and the 
LPA equation can be solved exactly. 

The paper is organized as follows: after introducing in Section \ref{sec2} the FRG treatment for 
the $O(N)$ model in dimension $d$, in Section \ref{sec3} we discuss for a general value of $N$
the occurrence of SSB in LPA when a finite number of couplings is used, we discuss the expansion
of the effective potential around the zero field; results for the Taylor expansion around the minimum
are also presented. The limit $N\to \infty$ 
is discussed in Section \ref{sec4}, while an LPA treatment of the appearance of SSB without 
truncations is presented in Section \ref{sec5}. Our conclusions are presented in Section \ref{sec6}, 
while in Appendix \ref{app:A} we collect useful informations on the regulator functions used in the 
main text and in Appendix \ref{app:B} we provide an alternative argument to show the validity 
of the Mermin-Wagner theorem at LPA level.

\section{Functional renormalization group for the $O(N)$ model}
\label{sec2} 
In the framework of the Kadanoff-Wilson \cite{Wi1971} RG approach the differential RG 
transformations are realized via a blocking construction consisting in the successive 
elimination of the degrees of freedom which lie above the running momentum cutoff $k$. 
Consequently, the effective theory defined by the blocked action contains quantum 
fluctuations whose frequencies are smaller than the momentum cutoff \cite{We1993,Mo1994}. 
This procedure generates the functional RG flow equation (Wetterich equation) for the 
effective action $\Gamma_k [\phi]$: 
\beq
k \partial_k \Gamma_k [\phi] = \hf \mathrm{Tr}  
\left[ k \partial_k R_k /(\Gamma^{(2)}_k [\phi] + R_k) \right],
\nonumber
\eeq 
where 
$\Gamma^{(2)}_k [\phi]$ denotes the second functional derivative of the effective action. 
$R_k$ is a properly chosen infrared (IR) regulator function which fulfills a few basic 
constraints to ensure that $\Gamma_k$ approaches the bare action in the UV limit 
($k\to\Lambda$) and the full quantum effective action in the IR limit ($k\to 0$): details are 
reported in  Appendix \ref{app:A}, where we also discuss the more commonly used 
regulators for $O(N)$ model and a more general choice able to recover all major types 
of regulators used in literature \cite{css}. Since RG equations are functional partial 
differential equations, it is not possible to solve them in general and approximations are 
required. The approximated RG flow depends on the choice of the regulator function $R$ 
and the physical results could become scheme-dependent. 

One of the commonly used systematic approximation is the truncated derivative expansion 
where the effective action is expanded in powers of the derivative of the field
\beq
\Gamma_k [\phi] = \int_x \left[U_k(\phi) 
+ Z_k(\phi) \hf (\partial_{\mu} \phi)^2 + ... \right].  
\nonumber
\eeq 
In LPA, the higher derivative terms are being neglected and the wave-function 
renormalization is set equal to a constant, i.e. $Z_k \equiv 1$. The solution of the 
RG equations sometimes requires further approximations: e.g., in case of the 
$O(N)$ symmetric scalar field theory the potential can be expanded in powers of 
the field variable around zero (with a truncation at the power $N_{\rm CUT}$). 

By using the dimensionless potential, $u_k \equiv k^{-d} U_k$, and dimensionless 
variables, the Taylor expansion of the potential around zero reads as
\bea  
\label{ansats1}
u_{k}(\phi) = \sum_{n=1}^{N_{\rm CUT}} \frac{1}{(2n)!} \, g_{n}(k) \, \phi^{2n}.
\eea  
It is convenient to introduce a field variable $\rho = (1/2) \phi^2$ then the Taylor 
expanded potential reads as 
\bea  
\label{ansats2}
u_{k}(\rho) = \sum_{n=1}^{N_{\rm CUT}} \frac{1}{n!} \, \lambda_{n}(k) \, \rho^{n}.
\eea  
As we can see the scale-dependence is encoded in the dimensionless coupling 
constants, which are related to each other as $g_{n}(k)/(2n -1)!! = \lambda_{n}(k)$. 

In LPA one obtains the following flow equation for the effective potential for the 
$d$-dimensional $O(N)$ model
\bea
\label{FeE}
\partial_t u  &=&  (d-2)\rho u' - d u + \frac{(N-1) A_d}{1+u'} + \frac{A_d}{1+u'+2\rho u ''}, \nn
A_d &=& \frac{1}{2^{d+1}}\frac{1}{\pi^{d/2}}\frac{1}{\Gamma(d/2)}\frac{4}{d}:
\eea
in \eq{FeE} dimensionless variables have been used and $\partial_t = k\partial_k$, 
$u' = \partial_\rho u$, $\Gamma(x)$ is the gamma function and for the sake of simplicity 
here we applied the Litim regulator (as can be seen from equation \eq{opt_lim} with $b=1$).

\section{The truncated $O(N)$ model ($N< \infty$, $N_{\rm CUT} < \infty$)}
\label{sec3}

\subsection*{Truncation around the zero field}

Let us first demonstrate the existence of SSB in the expanded O(N) model with finite 
$N_{\rm CUT}$ and finite $N$. 

\begin{figure}[ht] 
\begin{center} 
\epsfig{file=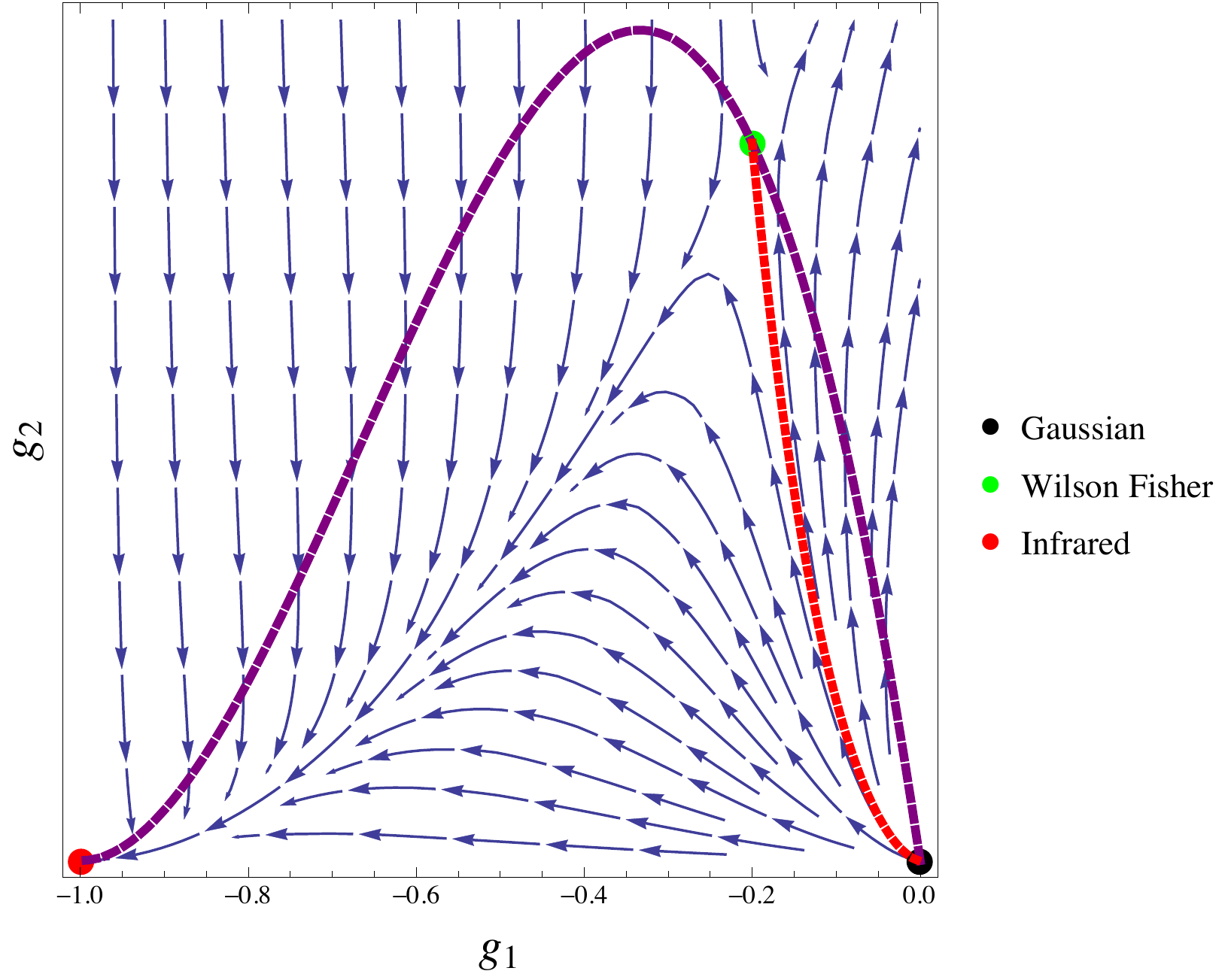,width=8.0 cm}
\caption{
\label{fig1}
Phase diagram of the $O(N=1)$ model for $d=1$ dimensions obtained by the numerical 
solution of the RG equations for two dimensionless couplings ($N_{\rm CUT}=2$) using the 
Litim regulator. Arrows indicate the direction of the flow. The red (dotted) line shows the 
separatrix and the purple (dashed) line stands for the vanishing mass beta function curve. 
The Gaussian (black), the Wilson-Fisher (green) and the IR convexity (red) fixed points are 
also shown. 
} 
\end{center}
\end{figure}

Let consider to start with the simplest case: two couplings ($N_{\rm CUT} = 2$) for the Ising 
case ($N=1$) in $d=1$. The RG flow equations for the couplings can be derived from \eq{FeE} 
and reads in this case
\bea
\label{g1_g2}
\partial_t g_1 = -2 g_1 - \frac{1}{\pi} \frac{g_2}{(1+g_1)^2} \\  
\partial_t g_2 = -3 g_2 +\frac{6}{\pi} \frac{g_2^2}{(1+g_1)^3}
\eea
which is obtained by using the Litim regulator. Similar equations are obtained for general $N$, 
$d$ and $N_{\rm CUT}$ (not reported here). In \fig{fig1} the RG flow diagram obtained from 
\eq{g1_g2} for the $1d$ $O(N=1)$ model with two couplings ($N_{\rm CUT} = 2$) for $d=1$. 
The model does not have any phase transition at finite temperature \cite{Mussardo10}, 
however a Wilson-Fisher (WF) fixed point is clearly visible in \fig{fig1}. 
%
%

The important point we want to stress is that flow diagrams similar to the one depicted in 
\fig{fig1} can be obtained for any finite $N$ and any finite $N_{\rm CUT}$ for dimensions 
$1 \leq d < 4$ (by using any regulator functions). In $d=4$, the WF fixed point (green dot in 
\fig{fig1}) merges to the Gaussian one (black). 
The IR fixed point (red) appears in any 
dimensions and is related to the convexity of the potential \cite{IR-1,IR-2,IR-3,IR-4,IR-5}. 

Although one finds similar flows for $d\leq 2$ and $d>2$, there is of course an important 
difference between the two cases. For $d\leq 2$the appearance of SSB is not allowed by 
the Mermin-Wagner theorem, but \fig{fig1} clearly signals the presence of SSB: the red curve 
from the Gaussian to the WF fixed points separates the phases and the RG trajectories run 
to the IR fixed point corresponds to the symmetry broken phase. (More comments on the 
$N=1$ will be given in the following.)

The WF fixed point is situated on the dashed purple line in \fig{fig1} which is determined by 
the vanishing mass beta function (to which we will refer as the VMB curve). Indeed, from 
\eq{g1_g2} one finds
\beq
g_2 = -2\pi g_1 (1+g_1)^2 
\eeq
which depends on $g_1$ and does not depend on higher order couplings even if
$N_{\rm CUT}$ is increased. As a consequence, the VMB curve on the $g_1, g_2$ 
plane remains unchanged for \textit{any} finite value of $N_{\rm CUT}$. Another important 
observation is that any fixed point should be situated on the VMB curve (by definition). 
The role of the VMB curve at LPA level was recently discussed in \cite{Codello14bis} in 
connection to the FRG determination of the central charge in $d=2$. 

\begin{figure}[ht] 
\begin{center} 
\epsfig{file=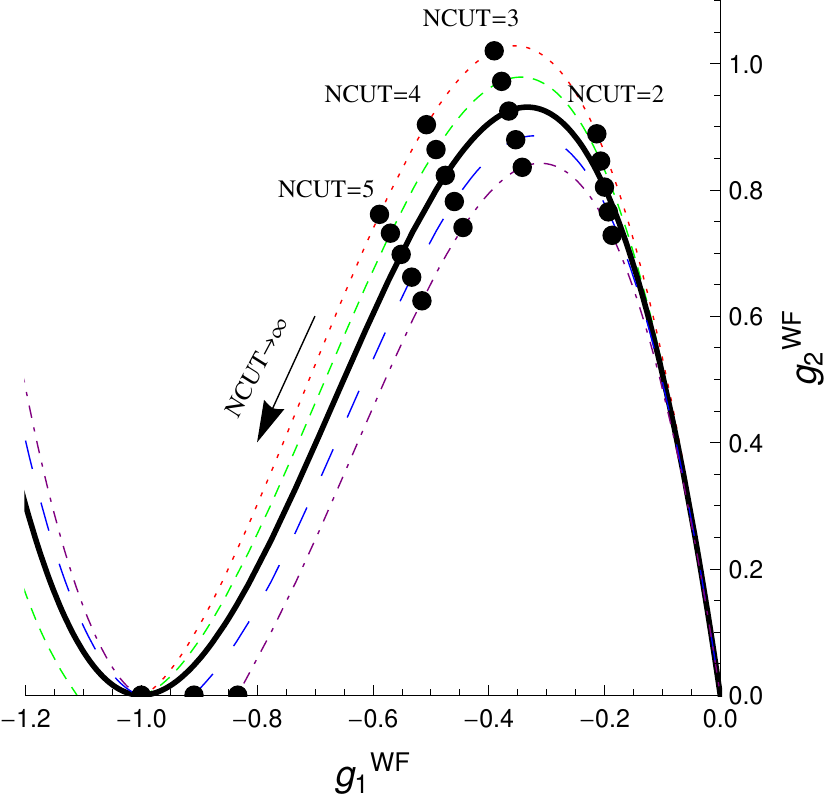,width=8.0 cm}
\caption{
\label{fig2}
Position of the WF fixed point on the VMB curves for the $O(N=1)$ model in $d=1$ 
for various values of $N_{\rm CUT}$. Different lines correspond to different regulators, i.e.
$0.8<h<1.2$ is chosen in \eq{litim_like}. The solid line corresponds to $h=1$, i.e. to the 
Litim regulator. The IR fixed point remains unchanged if $h \leq 1$.} 
\end{center}
\end{figure}

The position of the WF fixed point on the VMB curve depends on $N_{\rm CUT}$. Similar 
VMB curves can be drawn for any regulator function with the same properties, as shown in 
\fig{fig2} by projecting on the $(g_1,g_2)$-plane and applying the so called Litim-like regulator 
class. Using the notation of Appendix \ref{app:A}, this regulator class corresponds to various 
values of the parameter $h$ but with $b=1$ ($h=1$, $b=1$ is just the Litim regulator):
\beq
\label{litim_like}
r  = \left(\frac{1}{y} - h\right) \Theta(1-h y)
\eeq
(the regulator \eq{litim_like} is obtained by taking the $c \to 0$ limit in \eq{css_norm}). 
%
%
We observe that the position of the Gaussian fixed point is scheme-independent and thus 
the VMB curves always start form zero and go through the scheme-dependent IR fixed point.

\begin{figure}[ht] 
\begin{center} 
\epsfig{file=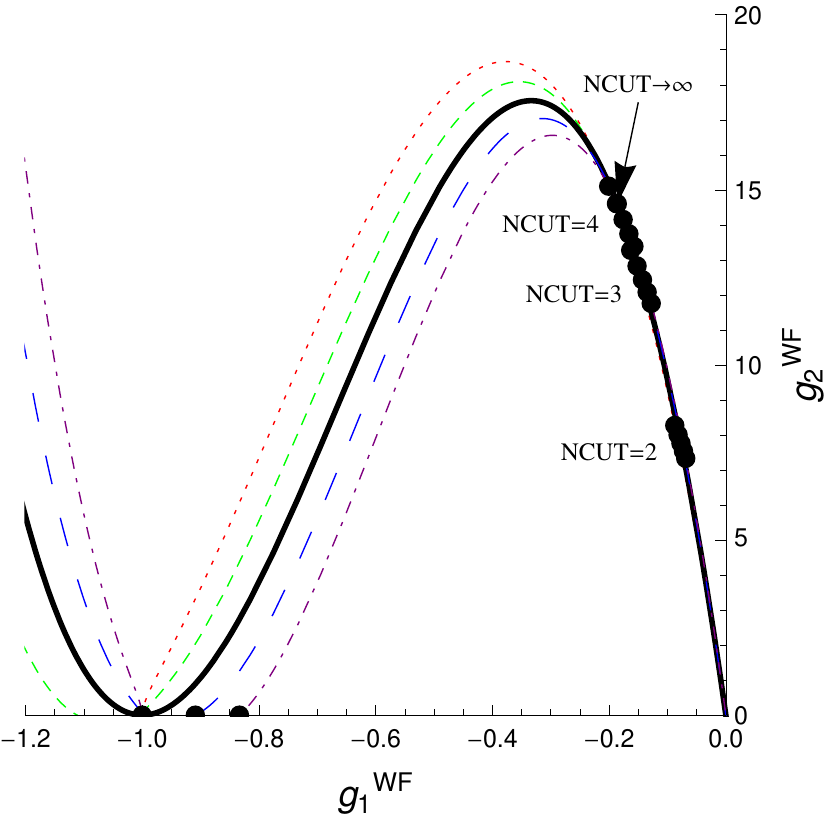,width=8.0 cm}
\caption{
\label{fig3}
Position of the WF fixed point on the VMB curves for the $O(N=1)$ model in $d=3$ for various 
values of $N_{\rm CUT}$. Different lines correspond to different regulators, i.e. $0.8<h<1.2$ is 
chosen in \eq{litim_like} as in \fig{fig2}. The $N_{\rm CUT}\to\infty$ WF fixed point (shown for 
the Litim regulator) is computed using the spike plot method \cite{Morris94,Codello12}.} 
\end{center}
\end{figure}

We found and verified that plots qualitatively similar to \fig{fig2} are obtained for general $N$ 
and $1\le d \leq 2$ (with d real): in these cases the WF fixed points corresponding to the different 
regulators for increasing $N_{\rm CUT}$ tend to the (respective) IR fixed points. This has to 
be contrasted with the situation in $d=3$ (or more generally for real $d\geq 2$), as it is shown 
in \fig{fig3}: namely, for increasing 
values of $N_{\rm CUT}$ the WF fixed points does not tend to the IR ones as for $d=1$, but 
tend to constant (non-trivial) WF fixed points for $d=3$, as indicated in \fig{fig3}. These 
non-trivial WF fixed points can be computed for $N_{\rm CUT} \to \infty$, i.e. treating exactly 
LPA, using the spike plot method \cite{Morris94,Codello12}. Clearly, the position of these 
non-trivial WF fixed points depend on the choice of the regulator.
%
%

\begin{figure}[ht] 
\begin{center} 
\epsfig{file=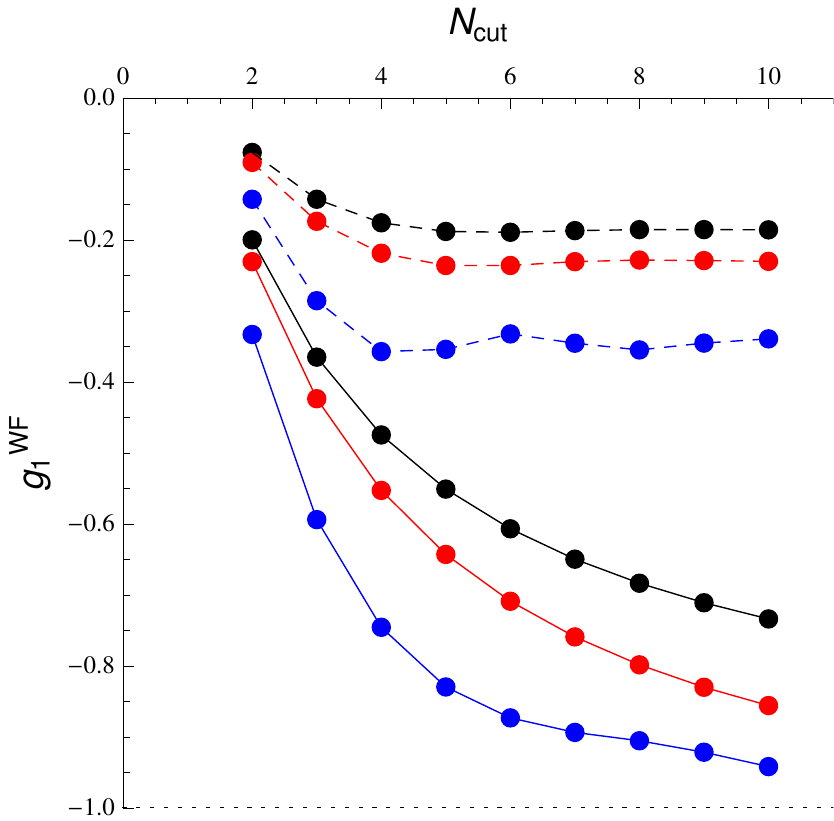,width=8.0 cm}
\caption{
\label{fig4}
The $g_1$ coordinate of the WF fixed point of the $O(N)$ model is shown as a function of 
$N_{\rm CUT}$ for $N=1$ (black), $N=2$ (red) and $N=10$ (blue) from top to bottom for 
$d=1$ (solid lines) and for $d=3$ (dashed lines).} 
\end{center}
\end{figure}

\begin{figure}[ht] 
\begin{center} 
\epsfig{file=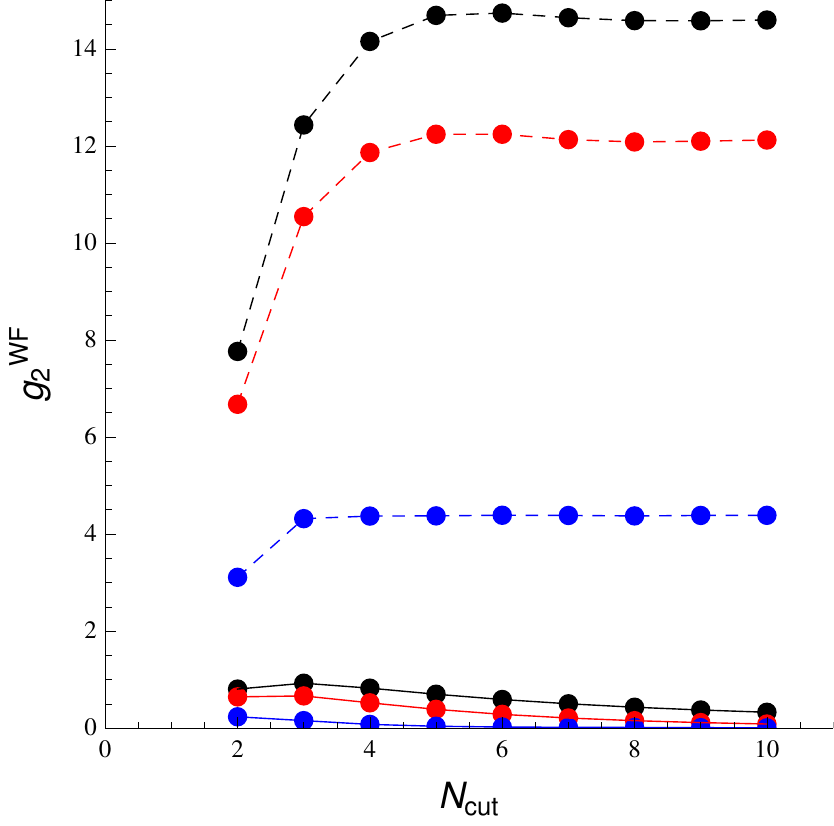,width=8.0 cm}
\caption{
\label{fig5}
The $g_2$ coordinate of the WF fixed point of the $O(N)$ model as a function of $N_{\rm CUT}$ : 
as in \fig{fig4} from top to bottom it is $N=1$ (black), $N=2$ (red) and $N=10$ (blue) for $d=1$ 
(solid lines) and for $d=3$ (dashed lines).} 
\end{center}
\end{figure}

Again, the scenario presented in \fig{fig3} is the same to what happens for general $N$ and 
$2 < d <4$. In order to illustrate this we plot the dependence of $g_1$ (\fig{fig4}) and $g_2$ 
(\fig{fig5}) on the truncation parameter $N_{\rm CUT}$ for $d=1$ and $d=3$ for three different 
values of $N$.
%
%
%
%
%

In summary, the results obtained for the $O(N)$ model with finite $N$ and finite $N_{\rm CUT}$ 
show the existence of SSB and a WF fixed point (distinct from the Gaussian one) for $1\leq d <4$ (with $d$ real), 
but indicated that for $N_{\rm CUT} \to \infty$ the symmetry broken phase disappears (persists) for 
$d\leq 2$ ($d>2$). This result has to be verified without operating the truncation of the couplings, 
i.e. treating exactly the LPA equation: we may refer to this as to the ``non-truncated'' $O(N)$  
model to emphasize the  fact that at LPA level no approximation is done. For finite $N$ this will be 
done in Section \ref{sec5}, while the next Section  is devoted to the spherical model limit $N \to \infty$.

\subsection*{Truncation around the minimum}
Now we will consider a Taylor expansion of the effective potential around the minimum,
\begin{equation}
u_{k}(\rho)=\sum_{i=2}^{N_{\rm CUT,m}}\frac{\lambda_{k,i}}{i!}(\rho-\rho_{0})^{i},
\end{equation}
where $N_{\rm CUT,m}$ is the truncation number around the minimum. In this case the results are 
drastically different. First of all we will consider this truncation at the minimum level $N_{\rm CUT,m}=2$. 
Obviously it is possible to relate the values of the coupling defined around the zero $g_{1}$ and $g_{2}$ 
with the values of the coupling $\lambda$ and the running minimum $\rho_{0}$. However
this relations, which give the correct result for the WF fixed point, are not working for the Gaussian 
fixed point, which is $g_{1}=g_{2}=0$, but no solution of the fixed point equations for $\rho_{0}$ and 
$\lambda$ has a vanishing $\rho_{0}$.\\
The truncation around the minimum includes just one running coupling $\lambda_{k,2}\equiv\lambda$ 
and the running minimum value $\rho_{0}$. From equation \eqref{FeE} we can obtain flow equations 
for these two quantities \cite{rg5}, we report them for general real dimension $d$ and $N$,
\begin{eqnarray}
\partial_{t} \rho_{0}=(d-2) \rho_{0}+A_{d}\left(1-N-\frac{3  }{(1 + 2 \rho_{0} \lambda)^2}\right)\\
\partial_{t}\lambda=\lambda\left(4-d-2A_{d}\left(N-1+\frac{18 \lambda}{(1 + 2 \rho_{0} \lambda)^3}\right)\right).
\end{eqnarray} 
One can look for the fixed point solutions for $\rho_{0}$ and $\lambda$.
In the particular case of the Ising model ($N=1$) one gets, for example, the Wilson-Fisher fixed point 
given by
\begin{eqnarray}
\label{WFTruncationARoundMinimum1}
\rho_{0}=\frac{4 (2 d-5)^{2} A_{d}}{3 (d-2)^{3}},\\ 
\label{WFTruncationARoundMinimum2}
\lambda= \frac{3 (4 - d) (d-2)^{3}}{16 (2d-5)^{3}A_{d}}.
\end{eqnarray}
The corresponding expression for $N>1$ are very lengthy and are not reported here.\\
From the solution \eqref{WFTruncationARoundMinimum1}--\eqref{WFTruncationARoundMinimum2} 
one sees that the value of the minimum $\rho_{0}$ is well defined (positive) for every value of $d$ 
as long as $d>2$: a similar result is valid for every $N$.\\ 
For $d>4$ the solution for $\lambda$ is negative, and, again, this is true for every $N$. Also it should 
be noted that the coupling $\lambda$, for $N=1$, is diverging at $d=2.5$ and turns out to be negative 
for $d<2.5$, thus giving an unphysical solution for $d<2.5$: we know that this is not true, since in $d=2$ 
for $N=1$ there is SSB. This is not valid in the general $N$ case where the coupling $\lambda$ has 
only a maximum and is not diverging at $d=2.5$ and no sign change is present at any value of $d>2$.\\
%
%
\begin{figure}[ht] 
\begin{center} 
\epsfig{file=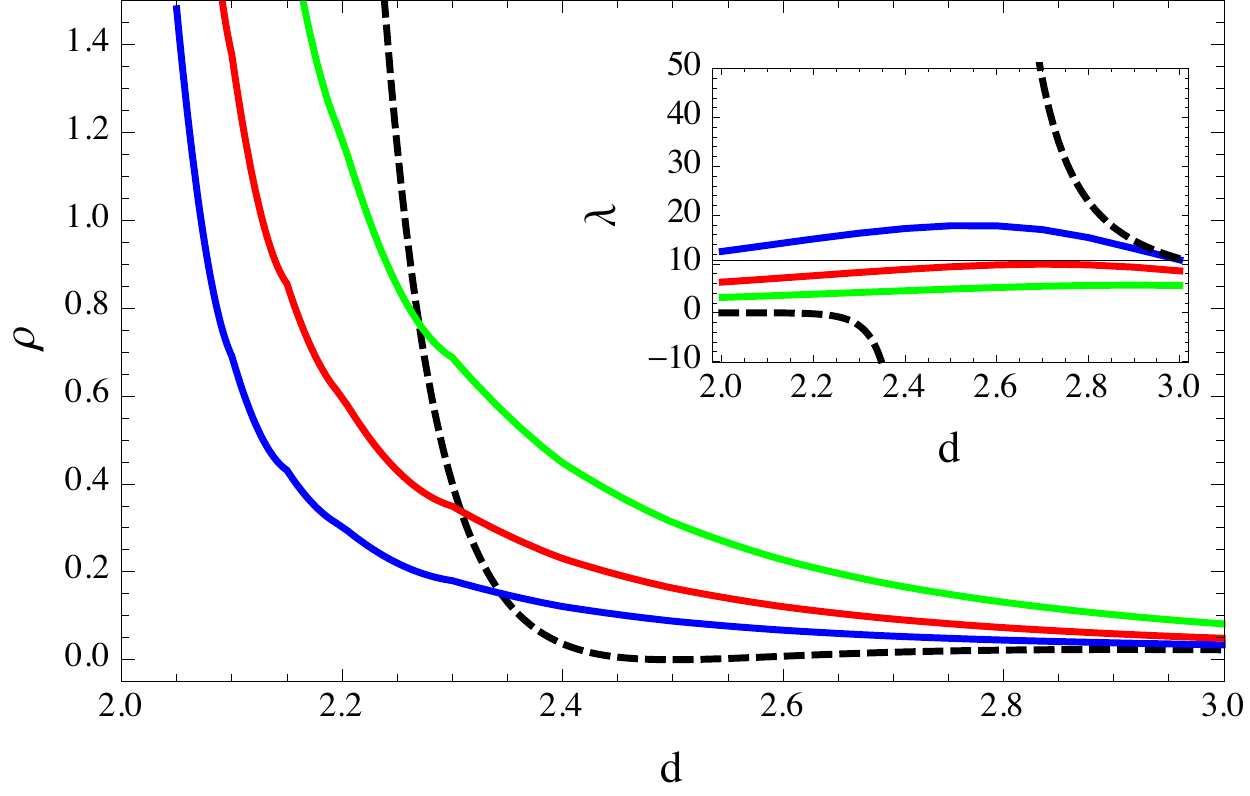,width=8.0 cm}
\caption{
\label{fig6}
Running minimum (main plot) and coupling (inset) values for the WF fixed point in the truncation around 
the minimum at $N_{\rm CUT,m}=2$ as a function of the dimension $d$, for the Ising model ($N=1$ black 
dashed curves), the XY model ($N=2$ , blue solid curves), the Heisenberg model ($N=3$, red solid curves) 
and the $N=5$ model (green solid curves). The Ising coupling $\lambda$ is the only one which is diverging 
and then turning negative at $d=2.5$ this is in contrast both with the well known exact solution of the Ising 
model in $d=2$ and with the following argument on exact solutions of equation \eqref{FeE}.
} 
\end{center}
\end{figure}
In \fig{fig6} is reported the minimum value $\rho_{0}$ as a function of the dimension at the WF fixed point for
various $N$ values, the minimum diverges for $d=2$ for every $N$ in agreement with the Mermin-Wagner
theorem. In the inset the values for the coupling $\lambda$ are shown. The coupling stays finite and positive 
for $d>2$ and $N>1$, but for $N=1$ it is diverging at $d=2.5$ and turning negative at $d<2.5$, then this 
truncation is not giving a reliable lower critical dimension for the Ising model.\\
We expect larger values of $N_{\rm CUT,m}$ do not to change the main qualitative results just presented.\\
This truncation, while giving the correct behavior for the SSB, cannot catch the non--interacting 
(Gaussian) fixed point and thus gives only a partial description of the theory phase space. 
In the next Sections we will show how it is possible to reproduce some of the correct result retrieved here 
and to go beyond them with a simple analysis of the exact flow equation for the effective potential \eqref{FeE}.

\section{The spherical model without truncations ($N = \infty$, $N_{\rm CUT}=\infty$)}
\label{sec4}
In this Section we consider the $O(N)$ model in the large $N$ limit: thus we can neglect the terms 
in \eq{FeE} which are in the order of $1/N$. To see this we are going to rescale \eq{FeE} by an 
irrelevant parameter $(A_d N)$, and considering the new variables $\rho\rightarrow \rho/(A_d N)$ 
and $u \rightarrow u/(A_d N)$. The derivative of the potential remains invariant under this rescaling 
$u'\rightarrow \frac{\partial u/(A_d N)}{\partial \rho/(A_d N)}=u'$. As a first step, we divide the RG 
\eq{FeE} by $A_d N$ keeping the potential non-truncated ($N_{\rm CUT}=\infty$): one finds
\begin{align}
\partial_t \frac{u}{A_d N}  = (d-2)\frac{\rho}{A_d N} u' - d \frac{u}{A_d N} + \frac{1}{1+u'}  \nonumber \\ 
-\frac{1}{N}\frac{1}{1+u'} + \frac{1}{N}\frac{1}{1+u'+2\rho u ''}.
\end{align}
Next we perform the rescaling
\bea
\partial_t u = (d-2)\rho u' - d u + \frac{1}{1+u'} -\frac{1}{N}\frac{1}{1+u'} \nonumber \\ 
+ \frac{1}{N}\frac{1}{1+u'+2\rho u ''}.
\eea
By taking the limit $N\rightarrow\infty$ the following terms remain:
\begin{equation}
\label{LaN}
\partial_t u = (d-2)\rho u' - d u + \frac{1}{1+u'}.
\end{equation}
This simplified expression represents the RG equation for the effective potential of the 
large $N$ O($N$) model in arbitrary dimension. From the equation \eq{LaN} we can 
extract some useful information. First we should differentiate it by $\rho$ in order to get 
an equation for the derivative of the potential. It reads then
\begin{eqnarray}\label{largeN}
\partial_t u' =(d-2) u' + (d-2) \rho u'' - d u' - \frac{u''}{(1+u')^2}.
\end{eqnarray}
Since in a physically reasonable theory the potential is bounded from below, we can assume 
that this potential has a global minimum at some $\rho=\rho_0$. For $\rho_0$ we have the following 
value for the derivatives of the potential at the fixed point: $u'(\rho_0)=0$, $u''(\rho_0)\equiv \lambda$. 
Assuming that the quartic coupling $\lambda$ is finite, we have then the following equation:
\begin{eqnarray}
0 = (d-2)\rho_0 \lambda - \lambda
\end{eqnarray}
with the solution
\begin{equation}
\label{MI}
\rho_0 =\frac{1}{d-2}
\end{equation}
which determines the cases where the minimum of the potential can be found or not in the 
large $N$ case. There is SSB if the potential has the minimum at some finite $\rho_0>0$: 
in the case of \eq{MI} we can satisfy this condition for $d>2$. For $d<2$ we find $\rho_0<0$, 
hence there will be no SSB. The $d=2$ case seems to be undefined, since $\rho_0\rightarrow\infty$ 
in this limit. However, if the minimum of the potential is sent to infinity one cannot define a proper 
minimum. The absence of a finite minimum indicates the absence of the spontaneous symmetry 
breaking for $d=2$ dimensions. This can be also seen by solving Eq.(\ref{largeN}) using the method 
of characteristics \cite{Pawlowski}. The large $N$ limit is a frequently used technique \cite{Edouard} 
where the results obtained can be considered as exact ones since the LPA approximation becomes 
exact when $N\rightarrow\infty$ \cite{zj}.

\section{The $O(N)$ model without truncations ($N<\infty$, $N_{\rm CUT} =\infty$)}
\label{sec5}
In this Section we finally consider the problem of determining the lower critical dimension for 
the $O(N)$ model for a finite $N$ but keeping the potential non-truncated in LPA. 

Let us first consider the $N=1$ case by trying the following strategy: numerically calculate the 
WF fixed point position for finite $N_{\rm CUT}$ and approximate the limit  $N_{\rm CUT}\to \infty$. 
Notice that, without knowing the exact WF fixed point positions, it is difficult for dimensions around 
$d=2$ to unambiguously extract from the limit of increasing $N_{\rm CUT}$ the value of the 
non-truncated model. For this reason we determine the WF fixed point for $N_{\rm CUT} =\infty$ 
by using the spike plot method \cite{Morris94,Codello12} in LPA.

The results are plotted in \fig{fig6} where we show that the $N_{\rm CUT}$ dependence of the 
WF fixed points on the VMB curves (obtained for Litim regulator) depends on the value of the 
dimension $d$: several $d$ between $d=1$ and $d=3$ (including $d=2$) are plotted for $N=1$. 
Similar plots are obtained for general value of $N$. The positions of the exact WF fixed points, 
computed by the spike plot method, are also indicated for each case (by the symbol $X$).

\fig{fig6} clearly shows that for $N_{\rm CUT} \to \infty$ the $g_2$ coordinate of the WF 
fixed point tends to a finite value for $d>2$ and runs to zero for $d\leq 2$: since this property 
is found to be valid in LPA for general values of $N$, when applied to $N  \ge 2$ this result 
implies that the LPA is enough to reproduce the content of the Mermin-Wagner theorem. 
For $d=2$ (sixth line from top in \fig{fig6}) one finds from the spike plot analysis that (for all $N$) 
no WF fixed point occurs: this result is correct for $N \ge 2$, but not for the Ising model ($N=1$).
As we will later comment, for the Ising model in $d=2$ one needs to apply LPA$'$ approximation. 
 
Given the clear numerical evidence that Mermin-Wagner is well obtained in the limit of increasing 
$N_{\rm CUT}$ and the excellent agreement with the spike plot method findings for the WF fixed 
points, we investigated and present in the following two analytical arguments  valid for the 
non-truncated (exactly treated) LPA confirming these results.
%
%
\begin{figure}[ht] 
\begin{center} 
\epsfig{file=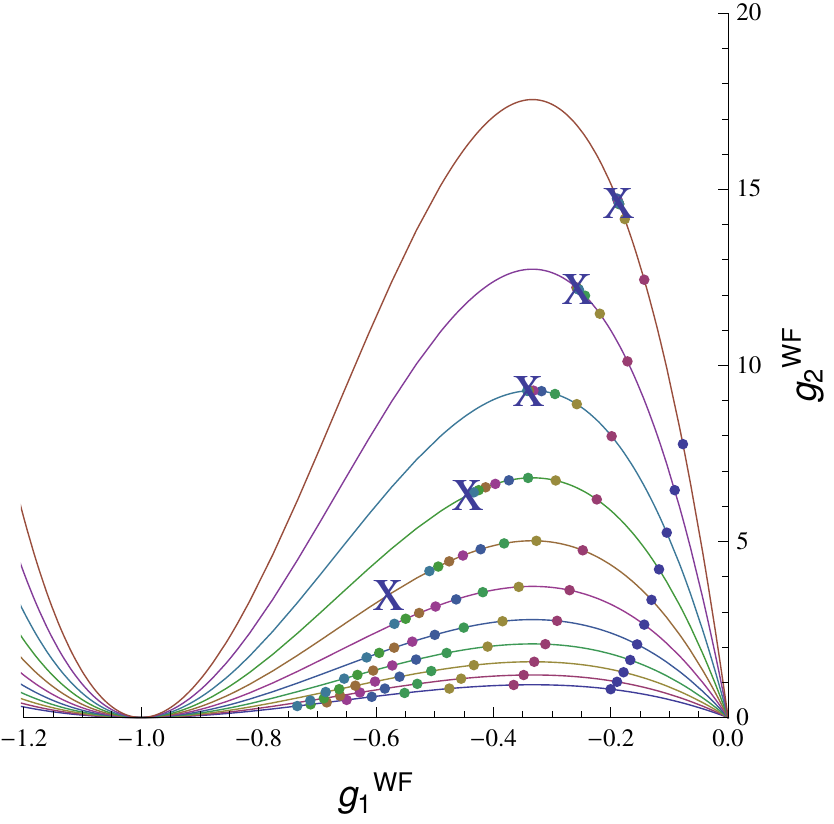,width=8.0 cm}
\caption{
\label{fig7}
The figure shows the positions of the WF fixed points and the corresponding VMB curves of 
the  $O(N=1)$ model for various values of $N_{\rm CUT}$ for $11$ different values of the 
dimension $1\leq d \leq 3$ having the values $d=3, 2.8, 2.6, 2.4, 2.2, 2, 1.8, 1.6, 1.4, 1.2, 1$. 
The VMB curves obtained for these dimensions are plotted from the top to the bottom in 
decreasing order. For each value of $d$, from right one has $N_{\rm CUT}=2,\cdots,10$. 
The exact WF points are indicated by the symbol $X$ and  are obtained by the spike plot 
method \cite{Morris94,Codello12}.} 
\end{center}
\end{figure}

In order to consider the appearance of SSB for the $O(N)$ model for finite $N$ but keeping 
the potential unexpanded ($N_{\rm CUT} =\infty$), let us start with the fixed point equation of 
\eq{FeE}: putting $\partial_tu=0$ one has 
\begin{equation}
\label{lpa_fixed}
d u - (d-2)\rho u' = \frac{A_d (N-1)}{1+u'}+\frac{A_d}{1+u'+2\rho u''}.
\end{equation}
The l.h.s of the RG equation \eq{lpa_fixed} is linear in the effective potential.
The r.h.s. depends on the effective potential and its derivatives non-linearly, thus we introduce 
the notation 
\begin{eqnarray}
LP &\equiv d u(\rho) - (d-2)\rho u'(\rho),\label{LP}\\
NLP &\equiv \frac{A_d (N-1)}{1+u'(\rho)} + \frac{A_d}{1+u'(\rho)+2\rho u''(\rho)}\label{NLP}.
\end{eqnarray}
where $LP$ ($NLP$) stands for the linear (non-linear) part. Let us consider the large field limit 
($\rho \gg 1$) of equation \eqref{lpa_fixed}. First of all we assume analyticity of the effective 
potential \cite{morris:96} at any finite value of the field, then the potential at infinity can only 
either be a constant or divergent. In the first case $NLP$ is constant at infinity and the potential 
will be just a constant at any $\rho$. In the second case $NLP$ in principle may vanish or tend 
to a constant (eventually zero) at infinity, thus the fixed point potential $u(\rho)$ must satisfy 
for large field the condition
\begin{equation}
\label{large_rho}
LP=C,
\end{equation}
where $C$ is a finite (or zero) constant to be consistently determined. Then one finds the 
solution for $\rho \gg 1$
\begin{equation}
u(\rho) = \frac{C}{d}+a \rho^{\frac{d}{d-2}},
\end{equation}
where $a$ is a proportionality constant. Now let us differentiate the previous expression of 
$u(\rho)$: this yields $u'(\rho)=a \frac{d}{d-2} \, \rho^{\frac{2}{d-2}}$, which for large $\rho$ gives 
a diverging quantity for $d>2$ and a zero for $d<2$. In the former case we are violating the 
assumption that $u'$ is bounded for large $\rho$, while in the latter the constant $C$ is zero. 
In both cases we find 
\begin{equation}
u(\rho)= a \rho^{\frac{d}{d-2}}
\end{equation}
for $\rho \rightarrow\infty$, 

The general solution of equation \eq{lpa_fixed}, which is not a constant, 
can be then divided into two parts,
\begin{equation}
\label{Effective_Potential_Solution}
u(\rho) = f(\rho)+a \rho^{\frac{d}{d-2}},
\end{equation}
where the function $f(\rho)$ is subjected to the condition 
\bea
\lim_{\rho\rightarrow\infty}f(\rho) = 0.
\nonumber
\eea
The physical Gibbs free energy $F(m)$ can be computed from 
the effective potential $u(\rho)$ 
passing from dimensionless variables to the dimensional ones 
\cite{wetterich:93}: one finds 
\begin{equation}
\label{Gibbs_Free_Energy}
F(m)=k^{d} u(k^{2-d}m^{2}) = k^{d}f(k^{2-d}m^{2})+a m^{\frac{2d}{d-2}},
\end{equation}
where $m$ is the dimensional field of our model, which in the case of a spin system is the 
average magnetic moment. The free energy of the system is obtained then in the $k\rightarrow 0$ 
limit of equation \eqref{Gibbs_Free_Energy}, where we should distinguish between three cases.

\subsection{$d>2$}
When $d>2$ the factor $k^{2-d}$ in the argument of the function $f(\rho)$, in 
eq.\eqref{Gibbs_Free_Energy} is diverging. However we know from the analysis of the 
general solution that the function $f(\rho)$ tends to a constant in the infinite limit of its 
argument, hence the Gibbs free energy for an $O(N)$ model in LPA for $d>2$ reads
\begin{equation}
\label{Fixed_Point_Free_Energy_High_d}
a m^{\frac{2d}{d-2}},
\end{equation}
where $a$ is positive and can be fixed following the procedure described in \cite{morris:96}.

\subsection{$d<2$}
In the case $d<2$ the situation drastically changes. Indeed, the factor $k^{2-d}$ in the 
argument of $f(\rho)$ in equation \eq{Gibbs_Free_Energy} is now vanishing. The behaviour 
of the $f(\rho)$ function for vanishing argument can be obtained from equation 
\eq{Effective_Potential_Solution}.

Since we know that such expression should be defined for any finite value of field 
$\rho$ \cite{morris:96}, then $f(\rho)$ is diverging in zero, in order to compensate the 
divergence of $\rho^{\frac{d}{d-2}}$, which has a negative exponent in $d<2$. Thus the 
behaviour of $f(\rho)$ in the limit of vanishing arguments is
\begin{equation}
\label{f_in_zero}
\lim_{x\rightarrow 0}f(x)=w(x)-a x^{\frac{d}{d-2}},
\end{equation}
where $w(x)$ is finite in zero. Substituting the last expression into eq.\eqref{Gibbs_Free_Energy} 
one obtains
\begin{equation}
\label{Fixed_Point_Free_Energy_Low_d}
k^{d}w(k^{2-d}m^{2}),
\end{equation}
which is zero in the $k\rightarrow 0$ limit. 

In summary, we obtained that for $d>2$ the critical free energy of a $O(N)$ model can have 
the form \eq{Fixed_Point_Free_Energy_High_d} or can be zero, thus the phase transition is 
present \cite{morris:96}. For $d<2$ the fixed point free energy can only be zero and no 
spontaneous symmetry breaking is allowed. In Appendix \ref{app:B} we report an alternative 
derivation of this result, again based on the analysis of LPA equation.

\subsection{$d=2$}
The previous argument cannot be directly used for $d=2$.  The numerical study of the 
equation \eqref{lpa_fixed}, i.e. $LP=NLP$, reveals in this case that for $N \ge 2$ the LPA 
gives the Mermin-Wagner result: indeed the solution for large field of $u(\rho)$ turns out to 
be oscillatory, therefore correctly implying the absence of SSB and the Mermin-Wagner theorem. 
However, such oscillatory solutions persist for $N=1$, predicting the absence of SSB, which is 
clearly wrong. A derivation of the fact that in LPA without truncations SSB does 
not occur in $d=2$ at LPA level is provided in Appendix \ref{app:B}.

To have a complete picture of the $d=2$ case one should go beyond LPA: as discussed 
in \cite{codello:13}, in LPA' the limit of the anomalous dimension $\eta$ for $d \to 2$ is 
vanishing provided that $N \ge 2$, and non-vanishing for $N=1$. This gives a clear 
explanation in the FRG framework of the presence or the lack of SSB in $O(N)$ models 
in $d=2$. Of course, the fact that there is no SSB for $N=2$ does not imply the absence 
of the Berezinskii Kosterlitz-Thouless transition \cite{ktb,ktb_1}, as can be seen also in FRG 
treatments \cite{frg_ktb_1,frg_ktb_2,frg_ktb_3,codello:13}.

\section{Conclusions}
\label{sec6}
In this paper we studied spontaneous symmetry breaking (SSB) for $O(N)$ models using 
functional renormalization group (FRG) techniques, showing that even the local potential 
approximation (LPA) when treated without further approximations is sufficient to give 
qualitatively correct results. For systems with continuous symmetry ($N \ge 2$) LPA gives 
no SSB for $d\le 2$ and SSB for $d>2$ in agreement with the Mermin-Wagner theorem and 
its extension to systems with fractional dimension; in particular, simple analytical expressions 
are found in the large $N$ limit, correctly retrieving the expected results for the spherical model. 
We observe that the presented results rule out any type of SSB, not only the standard (bicritical) 
Wilson-Fisher (WF) fixed point, but also all the other possible multicritical fixed points.

As a tool to assess the validity of different truncation schemes, for general $N$ we studied 
the solutions of the LPA renormalization group equations using a finite number of terms 
(and different regulators), showing that SSB always occurs even where it should not 
(i.e. for $d\le2$ for $N\ge2$). The SSB is signalled by WF fixed points which for any possible 
truncation are shown to stay on the line defined by vanishing mass beta functions. Increasing 
the number of couplings these WF fixed points tend to the infrared convexity fixed point for 
$d \le 2$ and to the pertinent exact LPA WF point for $d>2$.   
Moreover we studied the case of Taylor expansion of the effective potential around the minimum
$\rho_{0}$. Even when this expansion is truncated at lowest order $N_{\rm CUT,m}=2$, it is possible 
to retrieve the correct behavior for the Mermin-Wagner theorem, since $\rho_{0}$ is diverging when
$d\rightarrow 2$. However at this order the truncation around the minimum cannot provide the expected
behavior for the $N=1$ case, in fact the coupling $\lambda$ diverges at $d=2.5$ and becomes negative
below this threshold, even if it is well known that in the Ising model the SSB occurs even at $d=2$.

For the Ising model ($N=1$) the SSB is shown to occur for $d>2$ (as it should be), but not for 
$d=2$ (as it should not be).  At variance, finding the correct results for $d=2$ and $N=1$, as well as 
for the Ising model in $1<d<2$, requires to go beyond LPA  since the anomalous dimension cannot 
be neglected: in $d=2$ the LPA without truncations is sufficient to explain the absence of SSB for 
$N \ge 2$, but not to predict the presence of SSB for the Ising model. To have qualitatively correct 
results in $d=2$ valid for all $N$ anomalous dimension effects as introduced in LPA$'$ have to be 
considered. This has been recently shown in \cite{codello:13}, which shows how LPA$'$ is able to 
reproduce numerically the behaviour predicted by the Mermin-Wagner theorem for $d \to 2$ and 
$N \ge 2$ (with the anomalous dimension $\eta\to0$ and the correlation length exponent 
$\nu \to \infty$ \cite{Codello14bis}), and correctly predicting at the same time SSB and a finite 
anomalous dimension exponent for the Ising model. We extended these results showing that when 
the anomalous dimension vanishes then no SSB transition is possible in $d\leq 2$ (as it happens 
for the $O(N\ge2)$ models). Motivated by these findings, a study based on FRG of the Ising model 
in dimensions smaller than $2$ is in our opinion worthwhile of future work.

\section*{Acknowledgement}
We thank A. Codello, G. Gori, E. Marchais,  J. M. Pawlowski and D. Zappal\`{a} for discussions. 
We are also grateful to A. Rancon for very useful comments on the effects of the truncation around the minimum.
This research was supported by the European Union and the State of Hungary, co-financed by the 
European Social Fund in the framework of TAMOP-4.2.4.A/ 2-11/1-2012-0001 'National Excellence 
Program', by the T\'AMOP 4.2.1./B-09/1/KONV-2010-0007 project and also through the project 
Supercomputer, the national virtual lab (grant no.: TAMOP-4.2.2.C-11/1/KONV-2012-0010). Support 
from the CNR/MTA Italy-Hungary 2013-2015 Joint Project "Non-perturbative field theory and 
strongly correlated systems" is gratefully acknowledged.\\

\appendix 
\section{Regulator functions}
\label{app:A}
Regulator functions have already been discussed in the literature by introducing their dimensionless 
form
\beq
R_k( p) = p^2 r(y),
\hskip 0.5cm
y=p^2/k^2,
\eeq
where $r(y)$ is dimensionless. Various types of regulator functions can be chosen, but a more general 
choice is the so called CSS regulator \cite{css} which recovers all major types of regulators in its 
appropriate limits. By using a particular normalization \cite{css_sg,css_pms} it has the following form
\begin{align}
\label{css_norm}
r_{\mr{css}}^{\mr{norm}}(y) =& \;
\frac{\exp[\ln(2) c]-1}{\exp\left[\frac{\ln(2) c y^{b}}{1 -h y^{b}}\right] -1}  
\theta(1-h y^b), 
\end{align}
with the Heaviside step function $\theta(y)$ where the limits are
\begin{subequations}
\label{css_norm_limits}
\begin{align}
\label{opt_lim}
\lim_{c\to 0,h\to 1} r_{\mr{css}}^{\mr{norm}} = & \;
\left(\frac{1}{y^b} -1\right) \theta(1-y^b), \\
\label{pow_lim}
\lim_{c\to 0, h \to 0} r_{\mr{css}}^{\mr{norm}} = & \;
\frac{1}{y^b}, \\ 
\label{exp_lim}
\lim_{c \to 1, h \to 0} r_{\mr{css}}^{\mr{norm}} = & \;
\frac{1}{\exp[\ln(2) y^b]-1}.
\end{align}
\end{subequations}
Thus, the CSS regulator has indeed the property to recover all major types of regulators: 
the Litim \cite{opt_rg}, the power-law \cite{Mo1994} and the exponential \cite{We1993} ones.

\section{Spontaneous symmetry breaking in local potential approximation at finite $N$}
\label{app:B}
In this Appendix we provide an alternative derivation of the validity of the Mermin-Wagner theorem 
in LPA. 

The fixed point equation \eq{lpa_fixed} for the potential, using the notation \eq{LP}-\eq{NLP}, reads 
$LP=NLP$. In the large field limit ($\rho\rightarrow \infty$) the potential could be diverging or 
bounded, hence tending to a constant value. Let us consider in detail the second case: $NLP$ is 
then converging to a constant, since $u'$ and $u''$ vanish and the following differential equation is 
found
\begin{equation}\label{diffe}
d u(\rho) - (d-2)\rho u'(\rho)=c.
\end{equation}
The solution for this equation is
\begin{equation}\label{diffe-sol}
u(\rho)=\frac{c}{d}+a\rho^{\frac{d}{d-2}}
\end{equation}
Here, $c$ is the constant representing the large field limit of the $NLP$ and $a$ is another constant, 
obtained from the integration. Considering our assumption on $u$ in the $\rho\rightarrow \infty$ limit, 
namely that it is a constant, the constant of the integration can take only one value: $a=0$. It follows 
that $u=c/d=\text{constant}$ (the asymptotics).

Let us now consider the case when $u(\rho)$ is diverging in the large field limit. In this case we need 
to distinguish three sub-cases considering the behaviour of the derivative, $u'(\rho)$, since it can be 
diverging, tending to a finite value or to zero. In the last two cases $NLP$ tends towards a constant 
again. So the differential equation which should be solved has the same form as \eq{diffe}. Hence the 
solution is again (\ref{diffe-sol}). In the case when $u'$ is also diverging $NLP$ tend to zero, hence the 
differential equation slightly modifies:
\begin{equation}
d u(\rho) - (d-2)\rho u'(\rho)=0
\end{equation}
yielding the solution:
\begin{equation}
u(\rho)=a\rho^{\frac{d}{d-2}}
\end{equation}
Now we can consider the constant of integration $a$ for each case. Due to the stability requirement 
of the potential, that is $u$ has to be bounded from below, $a$ is being forced to be a positive real for 
all the three sub-cases. We can then write the form of the potential in the following as
\begin{equation}\label{fullro}
u(\rho)=g(\rho)+a\rho^{\frac{d}{d-2}}
\end{equation} 
where, $g(\rho)$ is a constant (or vanishes) in the large $\rho$ limit.

We are looking for the minimum $\rho_0$: let us differentiate equation \eq{fullro} and take it at 
$\rho=\rho_0$, which is assumed to be the minimum. Performing this operation one gets
\begin{equation}\label{deriv}
0=g'(\rho_0)+a \frac{d}{d-2}\rho_0^{\frac{d}{d-2}-1}
\end{equation}
The minimum can be then expressed as 
\begin{equation}\label{mini}
\rho_0=\left(-\frac{g'(\rho_0)}{a}\right)^{\frac{d-2}{2}}\left(\frac{d-2}{d} \right)^{\frac{d-2}{2}}
\end{equation}
We can now distinguish three sub-cases:
\begin{itemize}
\item{} 
for $d>2$ the second factor in the expression of the minimum has a positive real value. 
We have established already that the in the first factor the denominator $a$ is positive. Therefore 
$g'(\rho_0)$ must be negative or zero in order to fulfill the equation \eq{deriv}. Hence altogether the 
fraction in the bracket must be positive independently from the dimension. So we found that for 
$d>2$ the $\rho_0$ can be either vanishing or finite positive. This indicates the presence of SSB.
\item{} 
for $d=2$ the second factor gives a $0^0$, which is indeterminate, or alternatively one can define 
it as $1$ if we consider the $d=2$ case as a limit ($d\rightarrow 2$). In this instance what one can 
see already in the \eq{deriv} is that if we assumed for $\rho_0$ to be a positive real, then $g'(\rho_0)$ 
would be $-\infty$ to compensate the second term. Hence \eq{mini} is undefined or alternatively if we 
say $d\rightarrow 2$, then $\rho_0=\infty$, which means there is no finite positive minimum to consider, 
therefore no SSB occours in $d=2$ limit \cite{Pawlowski}.
\item{} 
for $d<2$ one can immediately see that the second factor in  \eq{mini} is going to have complex 
value(s). From equation \eq{deriv} we can conclude that $g'(\rho_0)\geq 0$ for $d<2$. The only value 
for $g'(\rho_0)$ that makes \ref{mini} physically sensible is when $g'(\rho_0)=0$, therefore the potential 
cannot have a true extremum (minimum) anywhere else than $\rho_0=0$. This clearly shows that there 
exists only a symmetric phase for dimensions $d<2$ in LPA.
\end{itemize}
For the $u\rightarrow \text{constant}$ case we can do essentially a similar argument. 
\par The conclusion is that the Mermin-Wagner theorem can be shown using FRG techniques in the LPA.


\end{document}